\documentclass[12pt]{iopart}
\newcommand{\Jpsi}{$J/\psi$ }
\newcommand{\pT}{$p_T$ }

\newcommand{\sNN}{$\sqrt{s_{\mathrm{NN}}}$ }
\newcommand{\pp}{$p+p$ }
\newcommand{\cucu}{Cu+Cu }
\newcommand{\auau}{Au+Au }
\newcommand{\raa}{$R_{AA}$ }
\usepackage{iopams}
\usepackage{graphicx}
\begin{document}

\title[\Jpsi production and correlation in \pp and \auau collisions at STAR]
{\Jpsi production and correlation in \pp and \auau collisions at
STAR}

\author{Zebo Tang (for the STAR collaboration)}

\address{Department of Modern Physics, University of Science and Technology of China,
Hefei, Anhui, China, 230026} \ead{zbtang@mail.ustc.edu.cn}

\begin{abstract}
The results on \Jpsi \pT spectra in 200 GeV \pp and \auau
collisions at STAR with \pT in the range of 3-10 GeV/$c$ are
presented. Nuclear modification factor of high-\pT \Jpsi is found
to be consistent with no suppression in peripheral \auau
collisions and significantly smaller than unity in central \auau
collisions. The \Jpsi elliptic flow is measured to be consistent
with no flow at $p_T < 10$ GeV/$c$ in 20-60\% \auau collisions.
\end{abstract}


\section{Introduction}
\Jpsi suppression in heavy-ion collisions due to color-screening
of its constituent quarks was proposed as the signature for the
formation of quark-gluon plasma by T. Masui and H. Satz 25 years
ago~\cite{colorscreen}. But results from SPS and RHIC showed some
other effects such as cold nuclear matter (CNM) effect and
recombination of charm quarks may play an important role in the
observed \Jpsi suppression in relativistic heavy-ion
collisions~\cite{Abreu:2000xe,PHENIX_Jpsi_AuAu}. It is believed
that high-\pT \Jpsi is less affected by CNM effect and charm quark
recombination effect, thus providing a cleaner probe to search for
evidence of color-screening effect in relativistic heavy-ion
collisions~\cite{Zhuang2009,XingboRalf2010,adscft}. STAR's
previous measurements showed no suppression for high-\pT \Jpsi in
\cucu collisions at 200 GeV, but with limited
statistics~\cite{starHighPtJpsiPaper}. And system size for \cucu
collisions may be too small. The same measurement in \auau
collisions with higher statistics is needed for better
understanding. On the other hand, the \Jpsi collective flow
measurement is crucial for the test of charm quark recombination
effect. It is also a clean probe to the charm quark flow in case
of coalescence hadronization.

The interpretation of \Jpsi modification by the medium created in
heavy-ion collisions also requires understanding of the quarkonium
production mechanism in hadronic collisions, but no model at
present fully explains the \Jpsi systematic observed in elementary
collisions. The \Jpsi spectrum measurement at intermediate and
high-\pT range and high-\pT $J/\psi$-hadron correlation
measurement may provide additional insights into the basic
processes underlying quarkonium production.

In this paper, we present the measurement of \Jpsi \pT spectra at
mid-rapidity with the STAR experiment in \pp and \auau collisions
at \sNN = 200 GeV in RHIC year 2009 and 2010 high luminosity runs.
We also present the measurement of \Jpsi elliptic flow $v_2$ from
low to high-\pT range in 20-60\% \auau collisions at \sNN = 200
GeV. The high-\pT $J/\psi$-hadron correlation in \pp collisions
has been discussed in Ref.~\cite{ZeboHP2010}.

\section{Results and Discussions}
The \Jpsi reconstruction method is similar to what we used in year
2005 and year 2006 data~\cite{starHighPtJpsiPaper,zeboThesis}. The
integrated luminosity used for this analysis is 1.8 $pb^{-1}$ (1.4
$nb^{-1}$) with transverse energy threshold $E_T > $ 2.6 (4.3) GeV
in \pp (\auau) collisions. Since year 2009, STAR installed a large
area Time-Of-Flight (TOF), consist of 72\% and 100\% full barrel
system at mid-rapidity ($|\eta|<0.9$) in year 2009 and 2010 run
respectively.
Including TOF in the analysis increases the signal-to-background
ratio of $J/\psi$ and reduces the statistical uncertainties.

\begin{figure}[th]
\begin{minipage}[c]{0.5\textwidth}
\centering
\includegraphics[width=0.75\textwidth,height=0.6\textwidth]{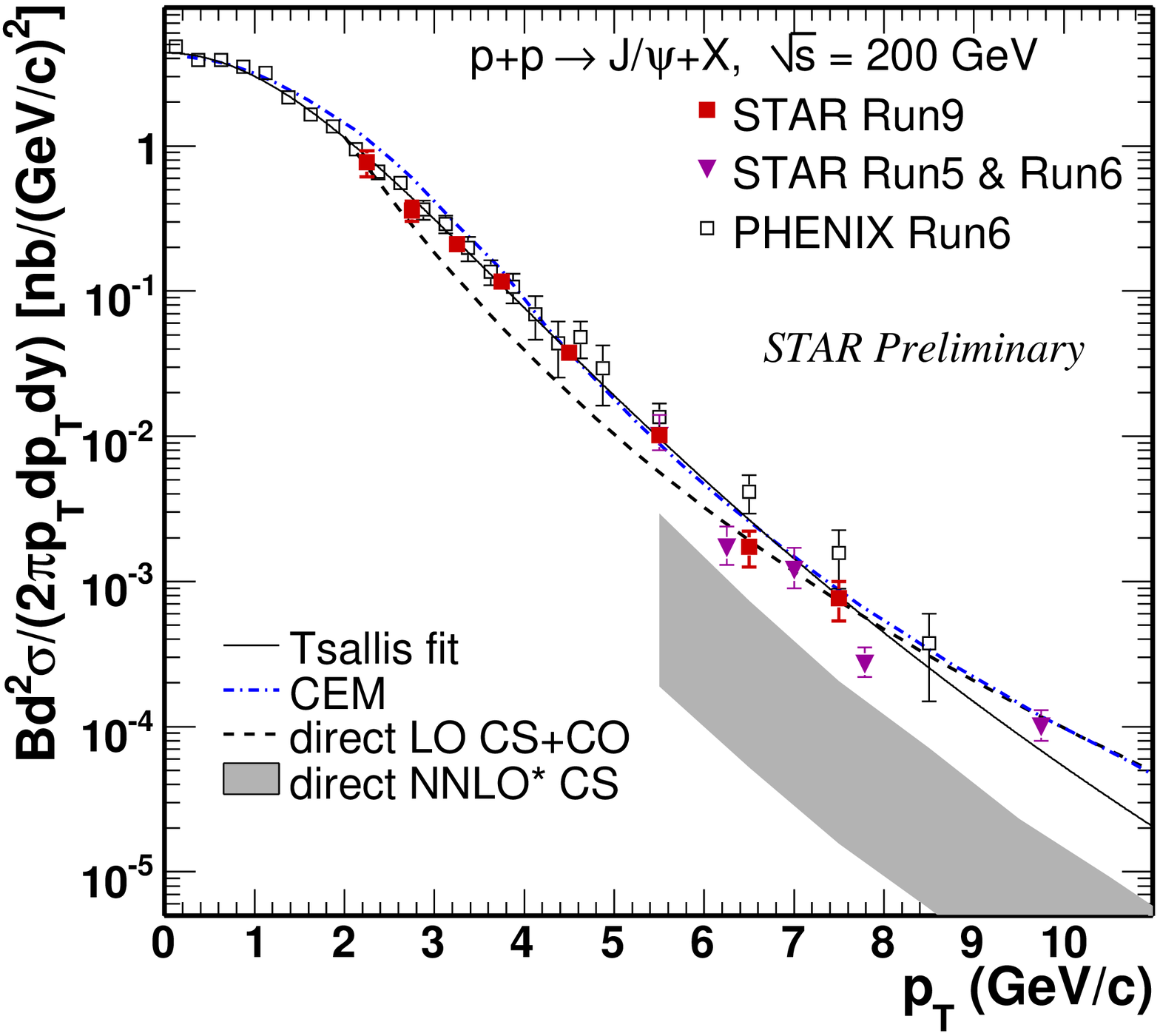}
\end{minipage}
\begin{minipage}[c]{0.5\textwidth}
\centering
\includegraphics[width=0.75\textwidth,height=0.6\textwidth]{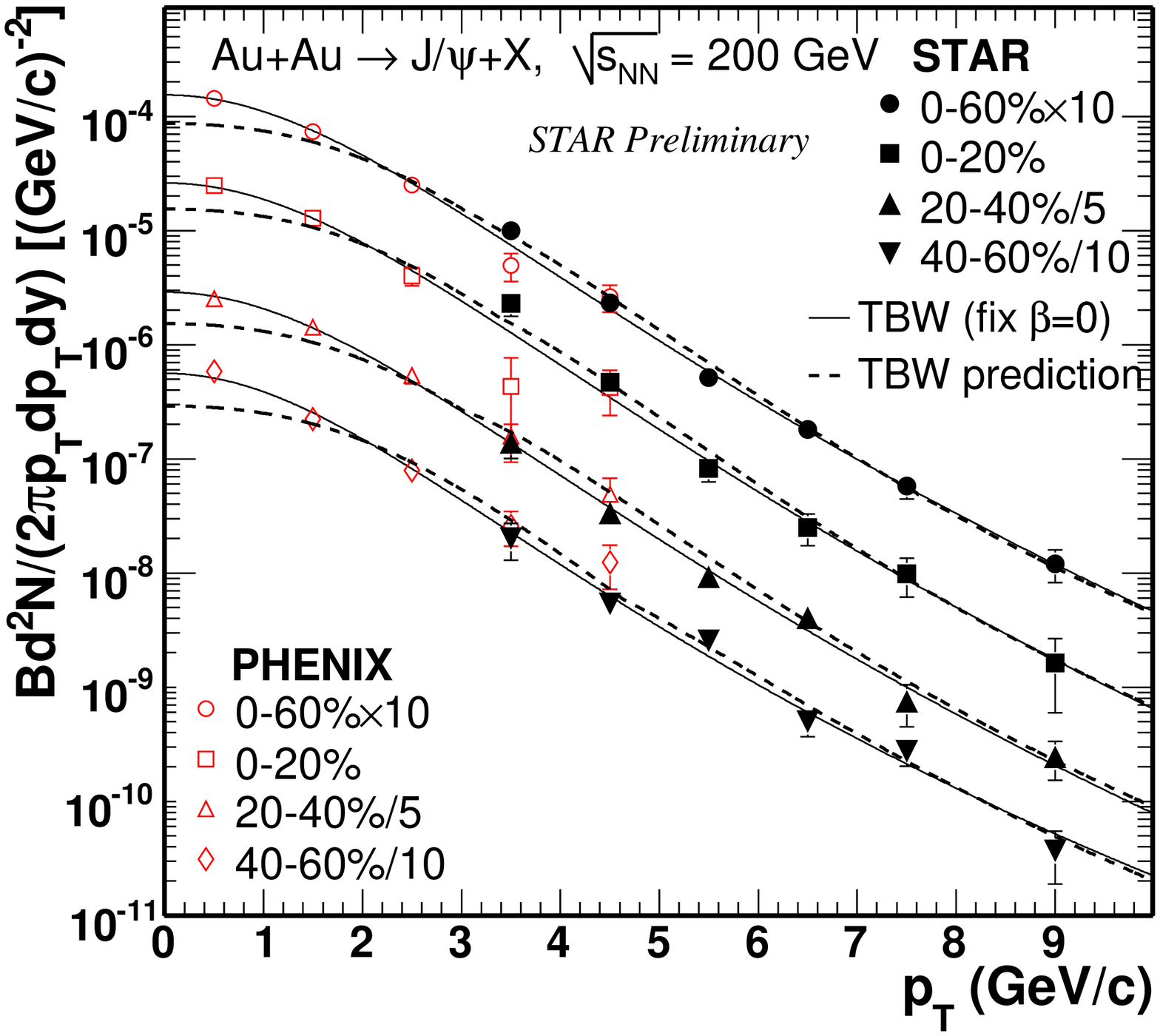}
\end{minipage}
\caption{\Jpsi \pT spectra in \pp (left) and \auau (right)
collisions.}\label{fig:spectra}
\end{figure}

The left panel of Fig.~\ref{fig:spectra} shows the fully corrected
\Jpsi \pT spectra in \pp collisions at $\sqrt{s}$ = 200 GeV. STAR
new measurements are consistent with previous STAR and PHENIX
measurements in the overlapping \pT region. The solid line
represents a Tsallis statistics based Blast-Wave (TBW) model
~\cite{Tsallis09,Tsallis11} fit to all of the data points. The
dashed line and gray band depict theoretical calculations of NRQCD
from color-coctet (CO) and color-singlet (CS)
transitions~\cite{Nayak:2003jp} and NNLO$^{\star}$ CS
result~\cite{Artoisenet:2008fc} for direct \Jpsi in \pp collisions
respectively. The CS+CO calculation leaves no room for feeddown
from $\psi'$, $\chi_c$ and $B$, estimated to be a factor of $\sim$
0.5 of direct $J/\psi$. NNLO$^{\star}$ CS predicts a steeper \pT
dependence. The dot-dashed line shows the calculation from color
evaporation model (CEM) for inclusive $J/\psi$, which can
reasonably well explain the \pT spectra at $p_T>$ 1
GeV/c~\cite{Bedjidian:2004gd}.

The right panel of Fig.~\ref{fig:spectra} shows the fully
corrected \Jpsi \pT spectra in \auau collisions with different
centralities. STAR and PHENIX measurements are consistent with
each other at the overlapped \pT range. The solid lines present
TBW fits to STAR and PHENIX data points simultaneously with radial
flow velocity $\beta$ fixed to 0, which can describe the data
points very well. The dashed lines show the TBW predictions
assuming \Jpsi has the same radial flow and freeze-out condition
as light hadrons, and they are much harder than the
measurements~\cite{Tsallis09,Tsallis11}. These indicate 1) \Jpsi
has very small (or 0) radial flow; and/or 2) there are significant
contribution from charm quark recombination at low $p_T$.

\begin{figure}[th]
\begin{minipage}[c]{0.5\textwidth}\centering\mbox{
\includegraphics[width=0.7\textwidth]{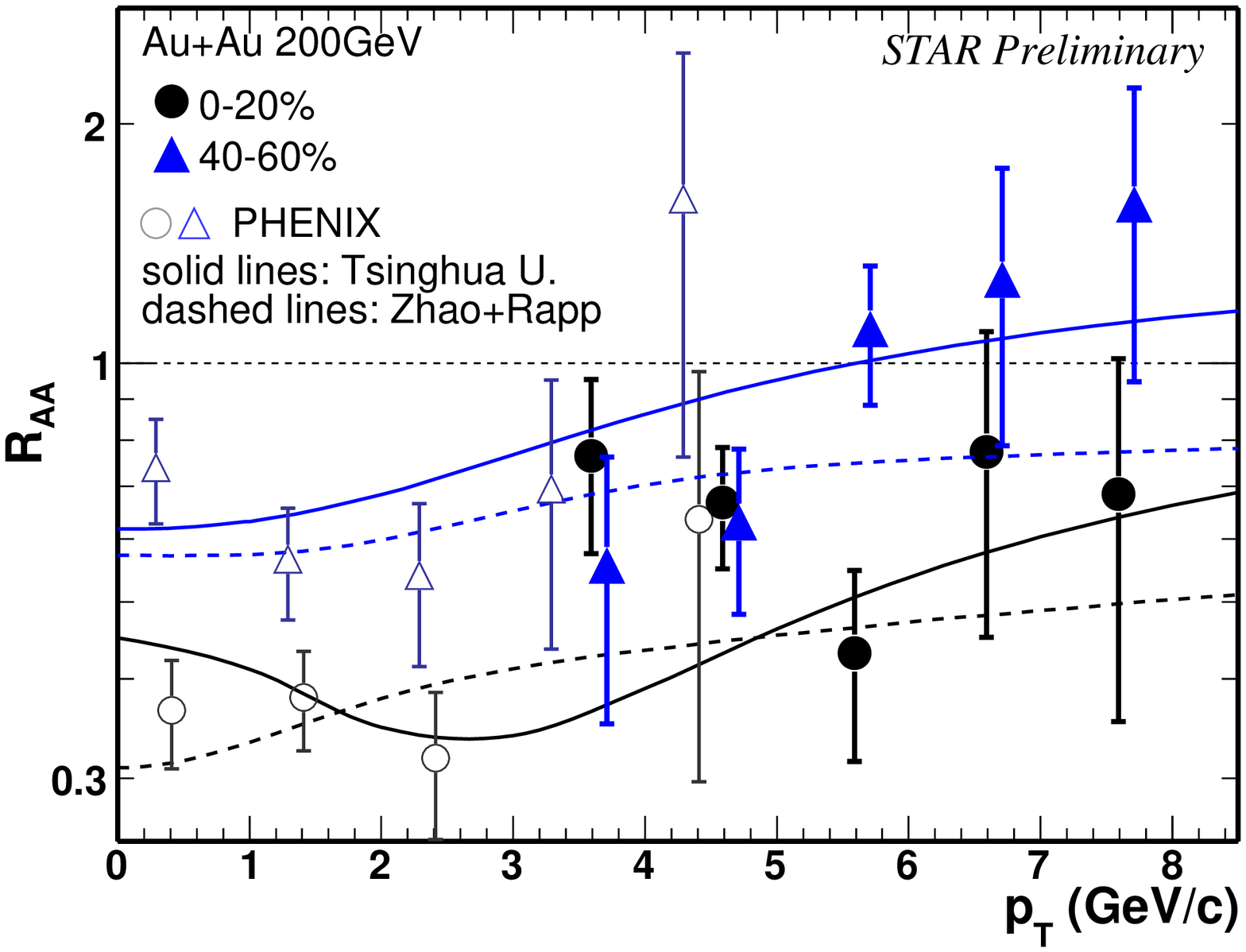}}
\end{minipage}
\begin{minipage}[c]{0.5\textwidth}\centering\mbox{
\includegraphics[width=0.7\textwidth]{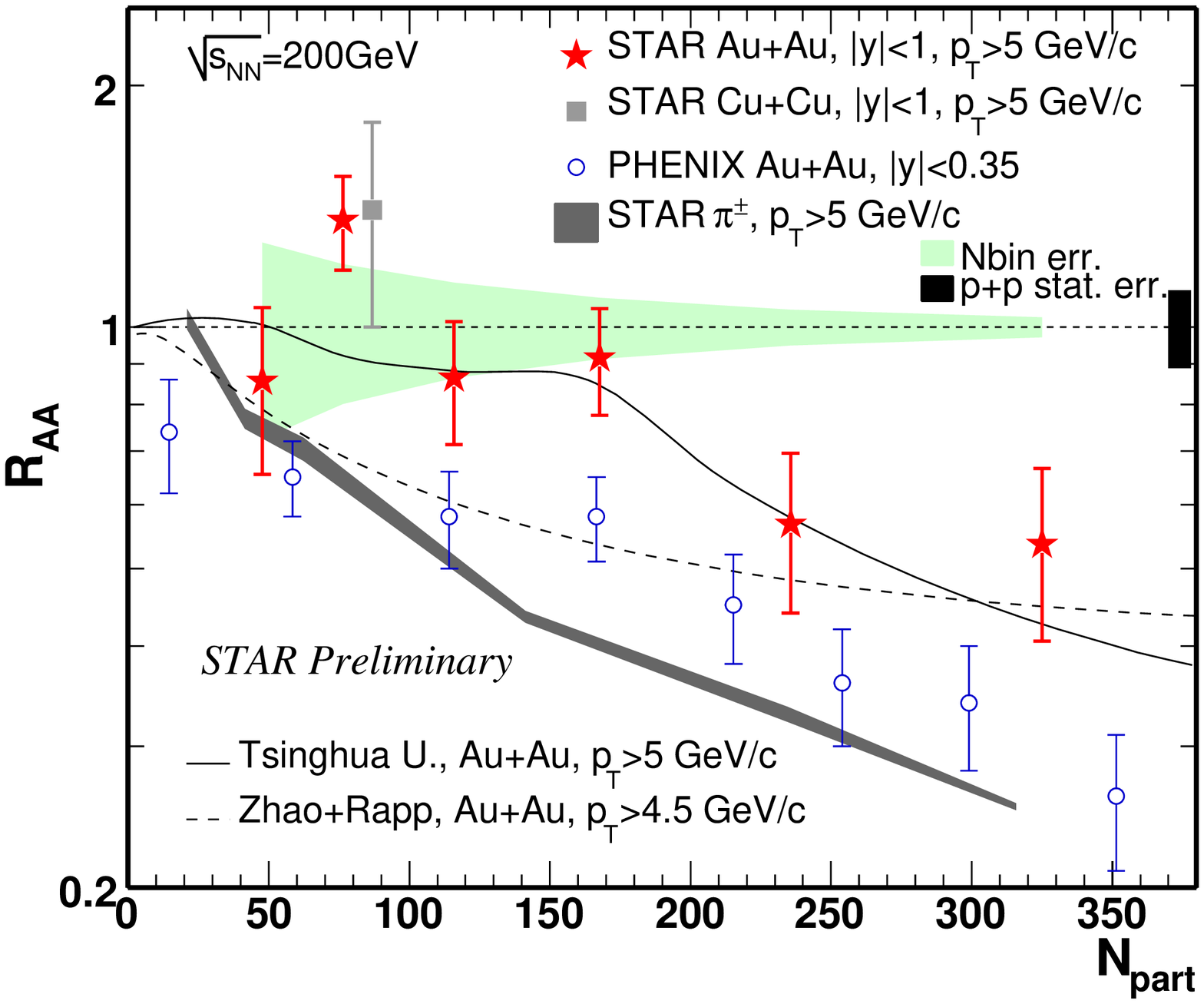}}
\end{minipage}
\caption{\textbf{Left:} \Jpsi \raa vs. \pT in 0-20\% and 40-60\%
\auau collisions. \textbf{Right:} \raa vs. $N_{part}$ for
low-$p_T$, high-\pT $J/\psi$s and high-\pT charged pion in \auau
collisions.} \label{fig:raa}
\end{figure}

The \Jpsi nuclear modification factor \raa as a function of \pT in
\auau collisions at different centralities measured by STAR at
high \pT are shown in the left panel of Fig.~\ref{fig:raa} and
compared to PHENIX measurements at low
$p_T$~\cite{PHENIX_Jpsi_AuAu}. There is a increasing trend from
low to high $p_T$, maybe due to CNM or \Jpsi formation time
effect. The high-\pT \Jpsi \raa is consistent with no suppression
in 40-60\% centrality, but systematically smaller than unity in
0-20\% centrality. The solid and dashed lines show two theoretical
calculations including both primordial \Jpsi and statistical charm
quark regeneration $J/\psi$~\cite{Zhuang2009,XingboRalf2010}.

The high-\pT ($p_T>5$ GeV/$c$) \Jpsi \raa as a function of number
of participants ($N_{part}$) in \auau collisions at \sNN = 200 GeV
are shown in the right panel of Fig.~\ref{fig:raa}. In peripheral
collisions (20-60\%), high-\pT \Jpsi has no suppression,
consistent with STAR previous measurements in \cucu collisions at
200 GeV. In central collisions (0-20\%), high-\pT \Jpsi is
significantly suppressed, which may be due to color-screening
effect. The \raa of low-\pT ($0<p_T<5$ GeV/$c$) \Jpsi measured by
PHENIX and high-\pT ($p_T>5$ GeV/$c$) charged pion measured by
STAR are also shown for comparison. The high-\pT \Jpsi \raa is
systematic higher than that for low-\pT $J/\psi$, and has
different trend from high-\pT charged pion.


\begin{figure}[th]
\begin{minipage}[c]{0.5\textwidth}
\centering
\includegraphics[width=0.8\textwidth]{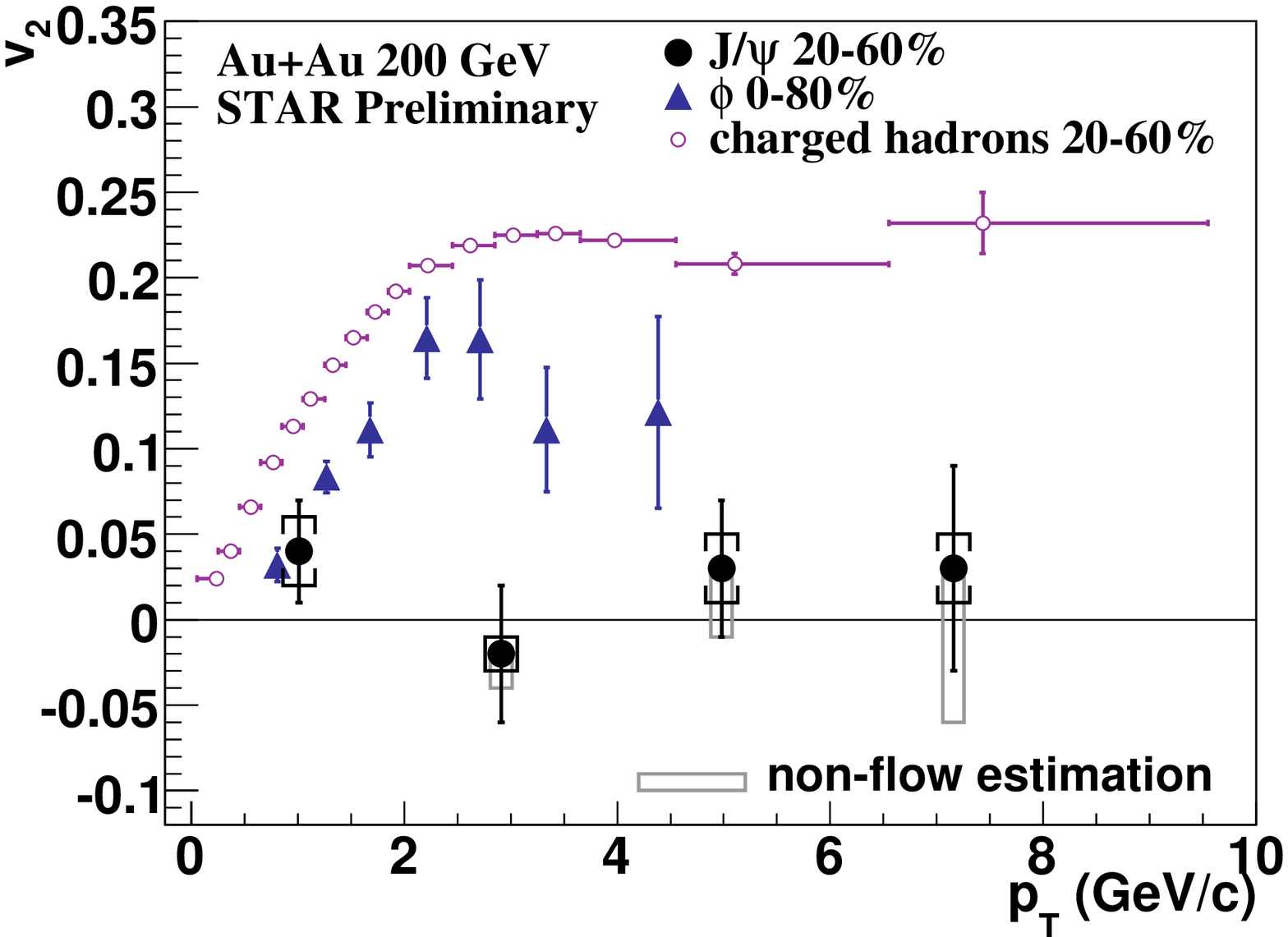}
\end{minipage}
\begin{minipage}[c]{0.5\textwidth}
\centering
\includegraphics[width=0.8\textwidth]{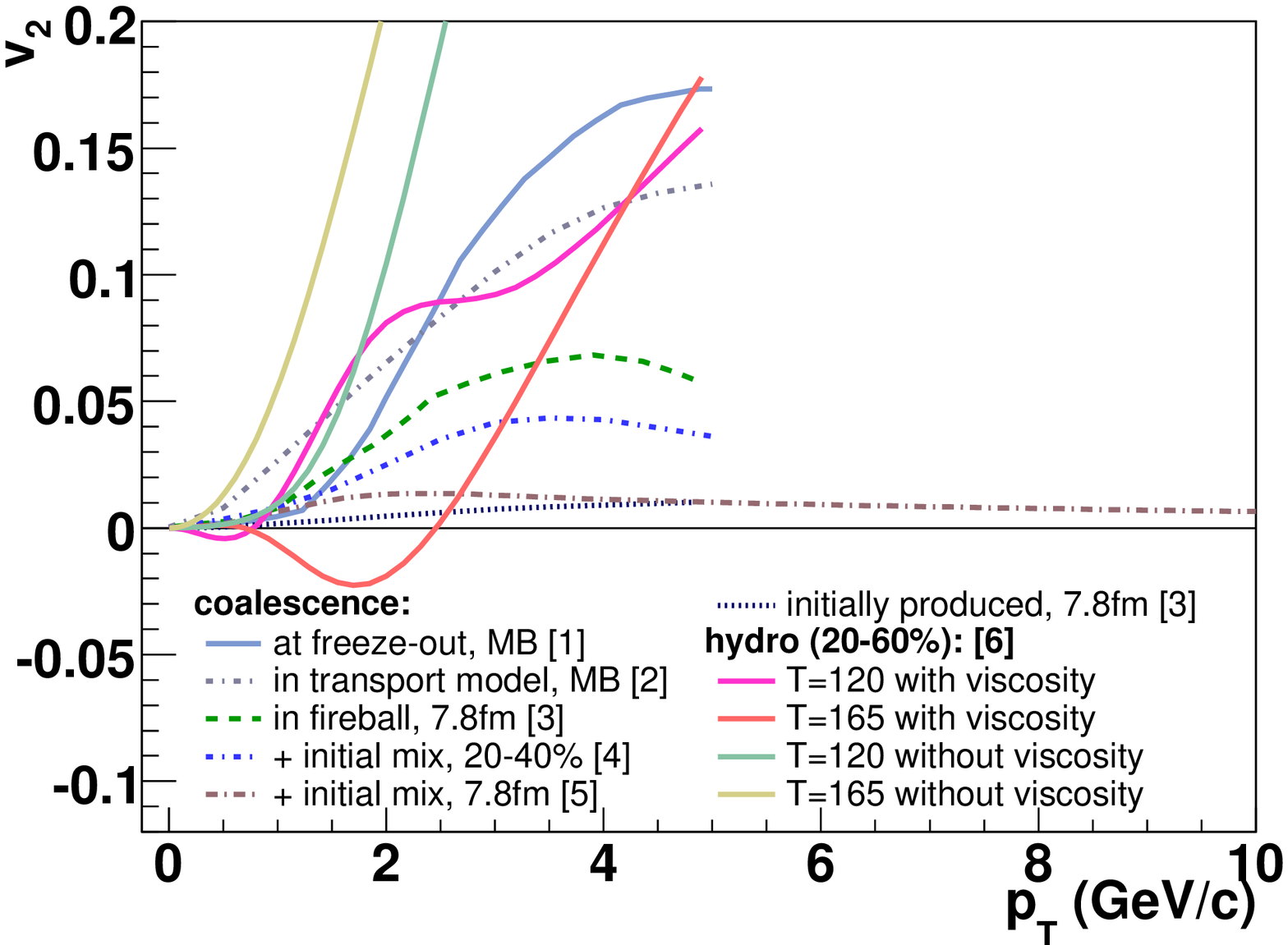}
\end{minipage}
\caption{\textbf{Left:} $v_2$ vs. \pT for $J/\psi$, $\phi$ and
charge hadrons in \auau collisions. \textbf{Right:} \Jpsi $v_2$
vs. \pT from model calculations, see text for detail.}
\label{fig:v2}
\end{figure}

For \Jpsi elliptic flow $v_2$ analysis~\cite{HaoPoster}, we use
all of the available data because this analysis does not need to
correct for absolute normalization and efficiency. The event
planes are reconstructed by using charged particles at
mid-rapidity measured by TPC. The $v_2$ results of inclusive
charged hadrons using these event planes are consistent with
previous STAR measurement. \Jpsi $v_2$ as a function of \pT in
20-60\% \auau collisions at \sNN = 200 GeV are shown in the left
panel of Fig.~\ref{fig:v2}, with the vertical lines (caps)
representing statistical (systematic) uncertainties, and the boxes
depict uncertainties from non-flow effect. \Jpsi $v_2$ is
consistent with zero in all of the measured \pT range within
uncertainties, and significantly lower than $\phi$ and inclusive
charged hadron $v_2$. Several model
calculations~\cite{Greco:2003vf,Ravagli:2007xx,Yan:2006ve,Zhao:2008vu,Liu:2009gx,UllrichHeinz}
with slightly different centralities are shown in the right panel
of Fig.~\ref{fig:v2}. The picture that \Jpsi production is
dominated by charm quark recombination with significant charm
quark flow is disfavored by STAR measurements, but models can
describe STAR measurements assuming that charm quark recombination
is dominant  at low \pT and primordial production is dominant at
high $p_T$.

\ack The author is supported in part by the National Natural
Science Foundation of China under Grant No. 11005103 and the China
Fundamental Research Funds for the Central Universities.
\section*{References}
\bibliography{QM2011_ZeboTang_v4}
\bibliographystyle{elsarticle-num}

\end{document}